\documentclass[conference]{IEEEtran}
\IEEEoverridecommandlockouts
\usepackage{balance}
\usepackage{cite}
\usepackage{amsmath,amssymb,amsfonts}
\usepackage{algorithmic}
\usepackage{graphicx}
\usepackage{textcomp}
\usepackage{xcolor}
\usepackage{comment}
\usepackage[export]{adjustbox}
\usepackage{multirow}

\def\BibTeX{{\rm B\kern-.05em{\sc i\kern-.025em b}\kern-.08em
    T\kern-.1667em\lower.7ex\hbox{E}\kern-.125emX}}
\begin{document}

\title{
Improving Sequential Recommender Systems with Online and In-store User Behavior
}

\author{
    \IEEEauthorblockN{Luyi Ma*, Aashika Padmanabhan*, Anjana Ganesh*, Shengwei Tang*,
    Jiao Chen, Xiaohan Li, Lalitesh Morishetti,\\ 
    Kaushiki Nag, Malay Patel, Jason Cho, Sushant Kumar, Kannan Achan}
    \IEEEauthorblockA{Walmart Global Tech, Personalization Team, Sunnyvale, CA, USA
    \\\{luyi.ma, aashika.padmanabhan,  anjana.ganesh, shengwei.tang, jiao.chen, xiaohan.li, lalitesh.morishetti, \\ kaushiki.nag, mpatel, jason.cho, sushant.kumar, kannan.achan\}@walmart.com}
    \thanks{\IEEEauthorrefmark{1}Both authors contributed equally to this research.}
}
% \author{Anonymous}

% \author{\IEEEauthorblockN{Anonymous.}
% \and
% \and
% \and
% \and
% \and

\maketitle

\begin{abstract}
%% abstract here
% The increasing demand for fashion recommendations in e-commerce platforms has posed challenges for outfit generation. 

Online e-commerce platforms have been extending in-store shopping, which allows users to keep the canonical online browsing and checkout experience while exploring in-store shopping. 
% In-store shopping becomes an extension of user online shopping interest because recently-browsed items could naturally inspires the user purchases in-store physically due to item availability and user purchase habit. Meanwhile, in-store shopping also reflects the fulfillment of user shopping interest, which is valuable for the online recommendation system to predict the future interaction.   
However, the growing transition between online and in-store becomes a challenge to online sequential recommender systems for future online interaction prediction due to the lack of holistic modeling of hybrid user behaviors (online \& in-store). The challenges are two-fold. 
First, combining online \& in-store user behavior data into a single data schema and supporting multiple stages in the model life cycle (pre-training, training, inference, etc.) organically needs a new data pipeline design. 
Second, online recommender systems , which solely relies on online user behavior sequences, must be redesigned to support online and in-store user data as input under the sequential modeling setting. 
% To overcome the first challenge, we propose a hybrid data pipeline to support online \& offline customer behavior data with the help of different level of information caching. 
To overcome the first challenge, we propose a hybrid, omnichannel data pipeline to compile online \& in-store user behavior data by caching information from diverse data sources.
Later, we introduce a model-agnostic encoder module to the sequential recommender system to interpret the user in-store transaction and augment the modeling capacity for better online interaction prediction given the hybrid user behavior. 
\end{abstract}


\section{Introduction}
\begin{figure}
    \centering
    \includegraphics[width=\linewidth]{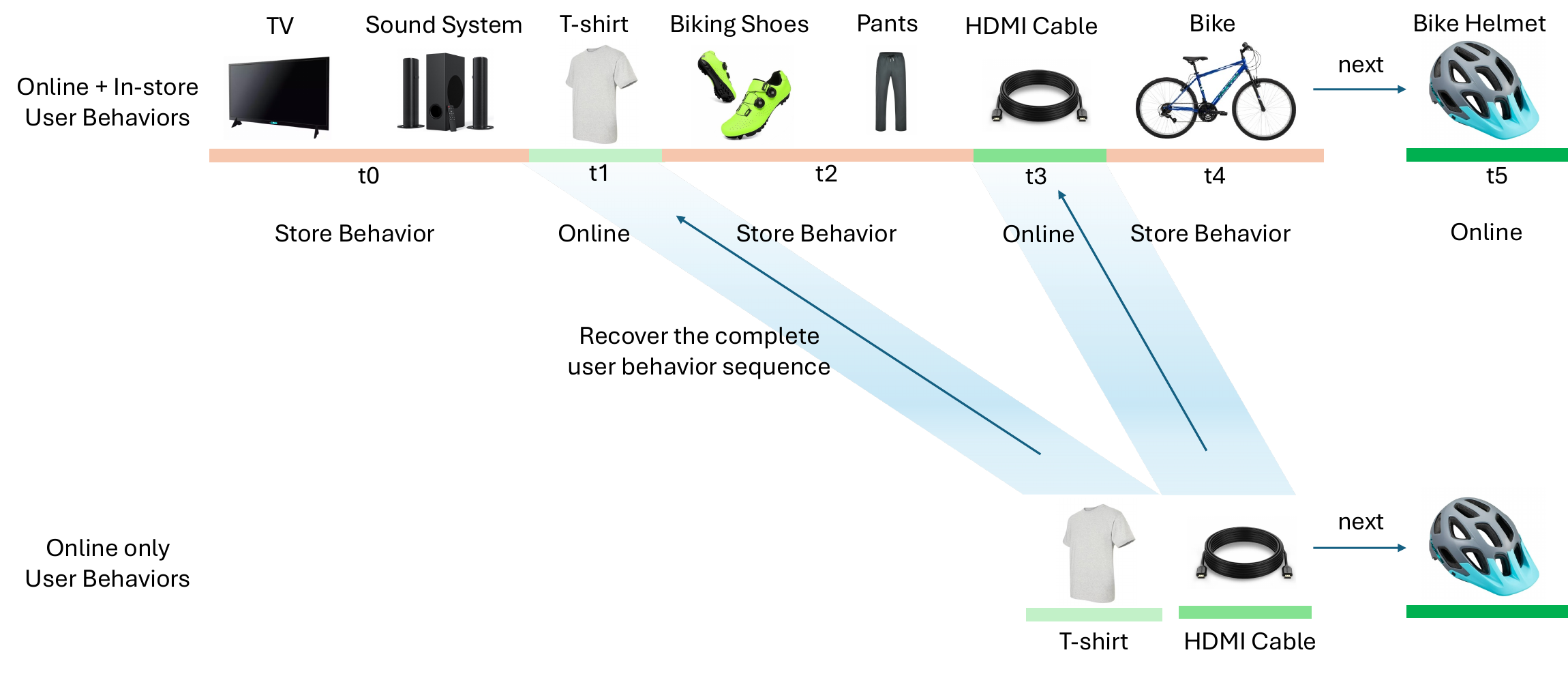}
    \caption{A user could have online and in-store shopping behaviors that correlate and reflect the user's shopping interest. Items in online user behavior are differentiable because of individual timestamps, while items from in-store user behaviors usually come in a set with shared timestamps. However, only online user behaviors can't unveil the complete shopping journey.}
    \label{fig:o2o-seqrec_example}
\end{figure}

Online e-commerce platforms have been extending in-store shopping which allows users to keep the canonical online browsing and checkout experience while exploring in-store shopping\footnote{https://www.bigcommerce.com/glossary/o2o-commerce/}. 
In-store shopping becomes an extension of user online shopping interest because recently-browsed items could naturally inspire the user to physically purchase in-store items due to item availability and user purchase habits. Meanwhile, in-store shopping also reflects the fulfillment of some user shopping interests, which is valuable for the online recommendation system to predict future interaction (see Figure~\ref{fig:o2o-seqrec_example}).

% \begin{figure*}
%     \centering
%     \includegraphics[width=\linewidth]{fig/Online-InStore.pdf}
%     \caption{A user could have online and in-store shopping behaviors that correlate and reflect the user's shopping interest. (a) Items in online user behavior are differentiable because of individual timestamps, while items from in-store user behaviors usually come in a set with shared timestamps. However, only online user behaviors can't unveil the complete shopping journey. (b) The transformer-based sequential recommender systems cannot process this hybrid sequence because of multiple items at the same timestamp, and we propose an encoder to enable the modeling of hybrid sequences. }
%     \label{fig:o2o-seqrec}
% \end{figure*}

However, leveraging in-store data to improve the online recommender systems for future online interaction prediction is challenging due to the lack of holistic modeling of hybrid user behaviors (online \& in-store).
The challenges of building such a recommender system are two-fold. 

First, combining online \& in-store user behavior data into a single data schema and supporting multiple stages in the model life cycle (pre-training, training, inference, etc.) organically needs a new data pipeline design. In-store transactions follow a different data schema than online transactions; due to the lack of information on user browsing, only the final transaction information is available, i.e., a set of items in the database with the same timestamp. Under the online shopping setting, users' browsing history are available in a streaming and real-time manner, and user-item interactions under different contexts (i.e., devices, pages, etc.) could be sorted by their timestamps to recover the journey. Thus, a hybrid data pipeline must combine two data sources regarding their different data update frequency, volume, and schema.

\begin{figure*}
    \centering
    \includegraphics[width=\linewidth]{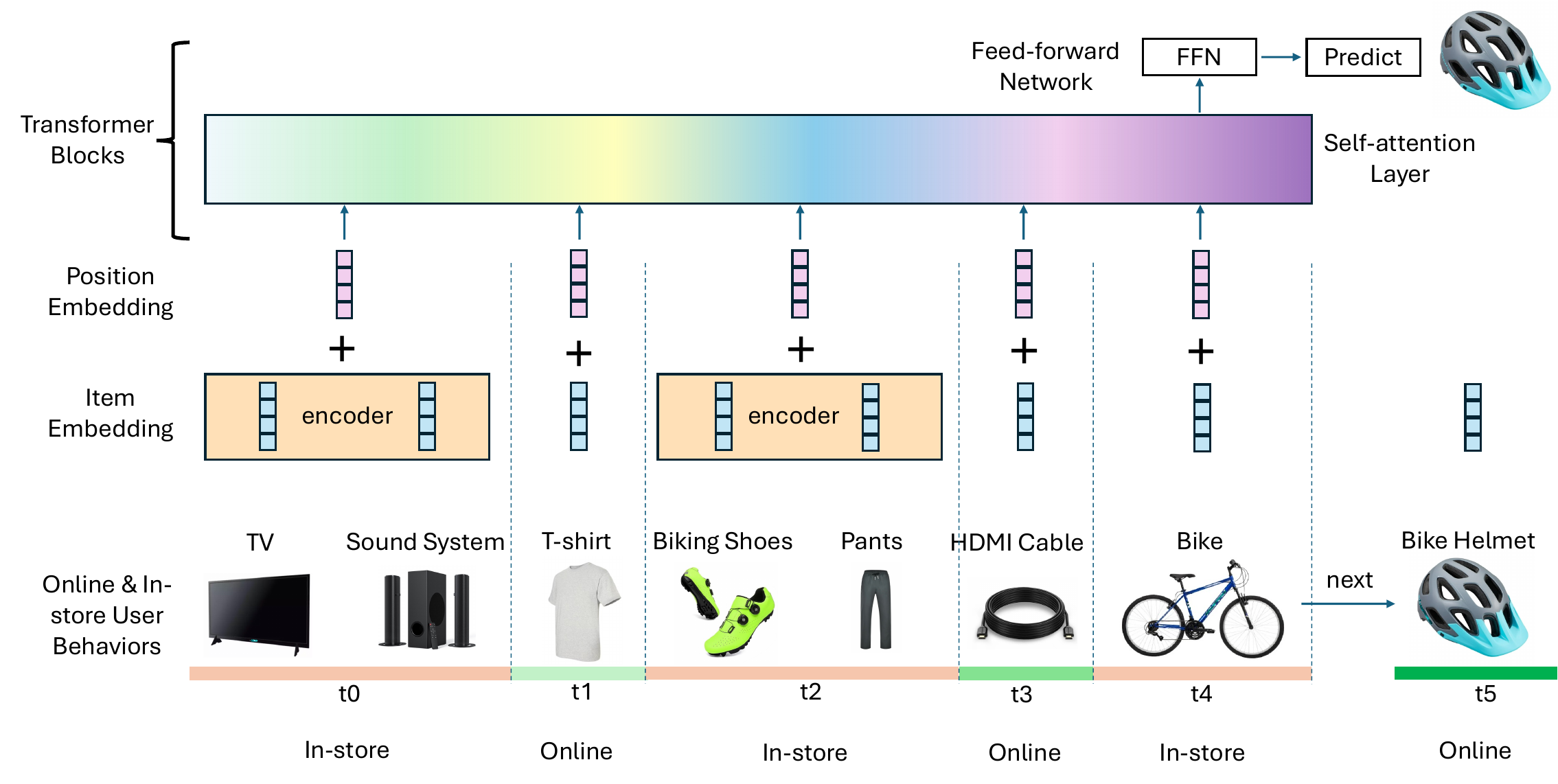}
    \caption{The transformer-based sequential recommender systems cannot process this hybrid sequence because of multiple items at the same timestamp, and we propose an encoder (light-orange box) to enable the modeling of hybrid sequences.}
    \label{fig:o2o-seqrec_transformer}
\end{figure*}

Second, the online recommender system, which relies solely on online user behavior sequences, must be redesigned to support online and in-store user data.
Modern online recommender systems follow Transformer-style modeling paradigms and mainly consider the online behavior signals~\cite{yuan2022multi, kang2018self}. Because user-item interactions can be sorted by timestamps, there is only one item interacted by the user at each timestamp. The prediction of the next behavior is based on the previous behavior sequence. Because of the sequence homogeneity and the self-regression characteristic, the Transformer-style design works well for the online shopping setting. 
When in-store shopping data are inserted into the online behavior sequence, the same practice is not optimal because the behavior sequences become heterogeneous, combining user-item set interactions and user-item interactions. 
Figure~\ref{fig:o2o-seqrec_transformer} illustrates this requirement on modeling by a decoder-only transformer design. 
Because the in-store data are item sets, multiple items map to a single timestamp, breaking the transformer backbone, which can only take one item at each timestamp. 

To overcome the first challenge of combining online and in-store data, we propose a hybrid pipeline by caching information from diverse data sources. To handle the massive volume of online user behaviors, a scalable fault-tolerant streaming processing system is considered for data processing and transformation. Aggregation happens through a configurable sliding time window to ensure that recent activities
are continuously factored in. In-store behaviors are collected at regular intervals. We design different feature registry mechanisms and caches to support the online and in-store user behaviors based on the downstream data use cases, such as model training, real-time inference, etc.
To tackle the second challenge of enabling in-store behavior to be incorporated into online user behavior sequence, we propose a design of the store transaction encoder, which could be a plug-in encoder module on Transformer-style sequential recommenders. 
We conduct experiments on a real-world dataset and demonstrate the benefits of combining online and in-store user data for modeling and the effectiveness of the store transaction encoder on general transformer-based sequential recommenders. 

Our contributions could be summarized as follows:
\begin{itemize}
    \item To our knowledge, we provide the first study on sequential recommender systems on online and in-store user behaviors. 
    \item We demonstrate that including in-store store behavior improves the performance of online sequential recommender systems, and the proposed store transaction encoder could further improve the performance. 
    \item We introduce a general design of the hybrid data pipeline for combining online and in-store user behaviors. 
\end{itemize}

\section{Related Work} \label{literature}
% outfit generation and recommendation
\subsection{Sequential Recommender Systems}
Sequential recommendation focuses on predicting users' preferences based on their interaction histories. Earlier studies have utilized various model architectures to encode these interaction sequences, such as Recurrent Neural Networks (RNNs)~\cite{DBLP:journals/corr/HidasiKBT15}, Convolutional Neural Networks (CNNs)~\cite{tang2018personalized} and Graph Neural Networks (GNNs)~\cite{wu2019session, li2020dynamic}. The recent success of Transformer~\cite{vaswani2017attention} has inspired the development of Transformer-based models for SR. SASRec~\cite{kang2018self} employs a unidirectional Transformer to dynamically assign weights to each interacted item. BERT4Rec~\cite{sun2019bert4rec} builds on this by utilizing a bidirectional Transformer with a Cloze task~\cite{devlin2018bert}, enabling the model to integrate user behavior from both the left and right contexts for each item. LSAN~\cite{li2021lightweight} improves upon SASRec by reducing model size, introducing a temporal context-aware embedding and a lightweight twin-attention network. ASReP~\cite{liu2021augmenting} addresses the data sparsity issue by using a pre-trained Transformer to enhance short user behavior sequences. Additionally,
Set Transformer~\cite{lee2019set} could also model heterogeneous sequences. Still, our prediction task is for the next singular item instead of a set, and the proposed solution is to enable general transformer-based SR for online \& in-store user behaviors. 

% [place holder]
% Sequential recommendation predicts the next item that a user interacts with given the user’s
% interaction history. Early attempts adopt probability models like
% Markov chains, while modern approaches adopt models like
% Recurrent Neural Networks (RNN), Graph Neural Networks
% (GNN), and Convolutional Neural Network (CNN) . More
% recently, due to its effective attention mechanism in sequential
% modeling tasks, Transformer-based recommendation
% models achieves best performance in sequential
% recommendation tasks. SASRec uses Transformer decoder as a
% sequential model, while BERT4Rec adopts bi-directional attention and masked token prediction as a training objective. However,
% conventional Transformer-based recommendation models do not incorporate the hybrid data type of online and in-store, which is crucial
% for understanding the intentions behind user interactions.

\subsection{Online and In-store User Behaviors and Item Recommendations}
In multiple previous studies, online shopping and in-store shopping behaviors have been reported to impact each other. Online shopping offers unparalleled convenience, a broader selection of products, and often lower prices, which appeal to a diverse range of consumers \cite{verhoef2015multi}. Several studies have reported that online shopping can substitute in-store shopping or reduce shopping trips in different cities and locations \cite{calderwood2014consumer, shi2019does}, especially during the COVID-19 pandemic period \cite{meister2023store}. 
% Online shopping behavior may also stimulates the in-store shopping behavior, for example, a customer travels to the store to get a product after finding it online in the nearby stores, or pick up an item purchased online. 
On the other hand, in-store shopping provides tactile experiences and immediate product availability, fostering customer satisfaction through sensory engagement and personal interaction \cite{grewal2020future, shamim2024s}, which can also impact customers' online shopping behavior. 
% Studies have reported that the accessibility and attractiveness of stores appear to have a negative effect on customers' online shopping behavior \cite{maat2018accessibility}. In addition, customers who has purchased a non-routine product in store are unlikely to purchase similar products in the near future.

However, recommender systems have been extensively explored using online and in-store data. Understanding these different shopping behaviors is crucial for predicting customers' future purchasing decisions in recommendation systems. Our work unveils the interplay between online and in-store shopping experiences and highlights the need for advanced predictive models that account for customer preferences in an omnichannel retail landscape.
%% preliminaries
\section{Preliminary}
This section defines the basic settings used in this paper and the research problem. 
Given a user $u$ from the entire user group $\mathcal{U}$, the online and in-store user behavior could be defined as a sequence of triplet $h_u = \{(i_1, t1, b_1), (s_1, t2, b_2), (i_2, t_3, b_3), ...\}$, where $i \in \mathcal{I}$ denotes an item, $s \subseteq \mathcal{I}$ is a set of items from the offline store behaviors and $b \in \{\textit{online}, \textit{in-store}\}$ indicates the data source of this record. 

In the traditional transformer-based sequential recommender systems, each item $i$ has its item embedding $v \in \mathcal{R}^{|d|}$ where $d$ denotes the embedding dimension. The transformer-based recommender systems are optimized to predict the next item of interest in the online shopping environment, given the previous sequence. 
We follow the same recommendation task on predicting the next item of interest in the online shopping environment to optimize the sequential recommender system given the hybrid user behavior sequence $h_u$, and study the impact of in-store user behaviors on predicting future online behaviors. 
\section{Data Pipeline} \label{sec:data_pipeline}
In this section, we will introduce our hybrid pipeline. 
The hybrid data pipeline is designed to integrate data on online and in-store customer behavior, addressing the limitations of existing sequential recommendation models that consider only online interactions. 
% As more and more customer interactions transition from online to in-store channels, a comprehensive understanding of customer behavior would require the combination of both real-time online data and batched in-store data. 
To achieve this, we propose a robust data pipeline that unifies multiple data sources into a single schema, enabling more accurate and insightful recommendations. 
In the rest of this section, we explain each pipeline component following the data flow. 

% \begin{figure*}
%     \centering
%     \includegraphics[width=0.8\linewidth]{fig/o2o-data-pipeline.png}
%     \caption{The hybrid data pipeline is designed to combine online and  (in-store) customer behavior data, merging various data sources into a unified schema. This integration improves both the accuracy and scalability of customer journey modeling. This approach not only streamlines the feature engineering process but also ensures that both historical and real-time data can be leveraged effectively for more informed decision-making.}
%     \label{fig:pipeline}
% \end{figure*}

\subsection{Data Sources and Data Ingestion}
This component is subdivided into two parts: online data ingestion and batch data ingestion. 
Online data are the streaming platform, including website visits, item views, clicks, search/browse patterns, etc.
% The continuous stream of real-time online data is crucial in understanding immediate customer preferences. 
Offline data refers to customer in-store data typically loaded in batches. This data is updated less frequently (e.g., daily, or hourly) and can come from a data warehouse or any other batch processing system. 
% This includes information on customer store transactions, in-store purchases, and returns. 
The data ingestion layer must ensure that data from online and offline sources is ingested into the pipeline with appropriate time synchronization.

\subsection{Data Processing and Transformation}
To process data collected from different sources, a streaming job (e.g., Spark Streaming\footnote{https://spark.apache.org/streaming/}, Flink\footnote{https://flink.apache.org/}) reads the real-time data from the online stream, aggregates the data and performs feature joins with offline sources and generates the latest customer features. 

% \noindent\textbf{Stream Aggregation}: 
The streaming job processes the real-time online data,  
% To generate significant time-based features, 
which processes
them in micro-batches at fixed time intervals, such as every 5 or 10 minutes. 
The configurable parameters like window duration and slide duration define the time interval and the shifting-forward step size. 
% how frequently the window shifts forward to recalculate the features, respectively. 
The streaming job joins the real-time data with offline in-store data on relevant customer or item attributes and ensure the alignment under different window durations. 
After the data is joined, additional transformations could be applied to generate the required features for the model. 
% The important aspect is to ensure that the online and offline data are aligned as they have different update frequencies and schema.
% For example, a 10-minute window with a 1-minute slide will aggregate customer interaction data every 10 minutes, updating every minute. 
% Aggregating over a certain window duration helps in capturing the cumulative behavior of customers, while the sliding duration will ensure that recent activities are continuously factored in. The overlapping windows ensure that no data is missed and that the customer features are updated in real time.

% \noindent\textbf{Join with Offline Data}: The streaming job joins the real-time data with offline in-store data on relevant customer attributes. The important aspect is to ensure that the online and offline data are aligned as they have different update frequencies and schemas.

% \noindent\textbf{Feature Engineering}: 
% These include calculating user preferences, product/brand affinity, engagement scores or other item-specific features. Furthermore, features are normalized, scaled or encoded as necessary for compatibility with the model.

\subsection{Feature Registry}
Next, the key features are registered in the feature registry, 
% which serves as a central repository. 
a critical component of the feature store and the hybrid data pipeline. 
It manages metadata about features including feature definitions, data types, and versions, which allows consistent use of features across both online and offline environments, and ensures that features are easily discoverable and reusable across different models. 

\noindent\textbf{Schema Management}: 
The feature registry helps manage and standardize the different schema of online and offline data sources. Online data typically arrive in real-time and may have different structures (e.g., session-based click stream data), whereas offline data often arrives in batches (e.g., in-store purchase history updated every hour or daily). The feature registry tracks feature definitions and ensures that when these data sources are combined, they adhere to a unified schema, which simplifies feature aggregation and joins. It ensures that both sources align in terms of the schema (e.g., timestamp alignment, feature name, etc.) and can be used together without conflicts.

\noindent\textbf{Version Control and Feature Validation}:
% In a hybrid pipeline, it is crucial to track the evolution of features as data sources and customer behaviors change over time. 
The feature registry supports version control, which allows tracking changes over time. This is essential for maintaining reproducibility in model training and inference and for enabling easy rollback. 
The registered features can also be reused across multiple models or pipelines if needed without re-engineering, which is efficient for modern machine learning models.  
% For example, in a cross-channel campaign personalization use case, features such as customer engagement score can be defined once and reused in both online marketing models and in-store promotion models. The registry ensures that the same feature is used uniformly, regardless of whether the data is coming from an online ad or from an in-store promotion.
Furthermore, it also supports feature validation, ensuring quality and correctness before features are deployed.
% While the feature engineering pipeline performs feature joins, such as merging real-time online customer actions with offline in-store purchases, the registry ensures that the features are appropriately mapped and compatible. The registry keeps track of the data type of the features and how they are related to defined entities 

\noindent\textbf{Seamless Data Transitions across Online and In-store}: 
% The feature registry also allows for seamless transitions between real-time and batch environments. It ensures that all registered features are available and are up-to-date for the relevant use cases. 
% In an omnichannel system, features that track online customer behavior such as recently browsed items, purchased products etc. and offline in-store spend data are essential to provide a seamless shopping experience. 
In an omnichannel system, user behavioral features should be accessible in real-time to offer dynamic pricing to the customer when they are browsing a product online or when they walk into a physical store. 
% For example, if a customer has shown interest in a product online but hasn't made a purchase, they might receive a discount when they visit the store, combining both their online browsing history and in-store purchase patterns. 
In such use cases, the feature registry ensures smooth interaction between the online and offline data by maintaining consistent, reusable, and versioned feature sets. 

\subsection{Online and Offline Stores}
% The feature store has offline and online data stores that support both batch processing for model training and low-latency access for inference.
Once the key features are created, they are stored in an online cache for low-latency access during model inference. They are also pushed to the offline data source (e.g., HDFS, BigQuery, cloud storage), making them reusable for model retraining without duplicate effort in feature engineering. The registered features can be used across multiple models and tasks, reducing redundancy in feature creation. This also ensures consistency on the feature definitions between training and inference pipelines, minimizing training-serving skew.
% By leveraging data streaming jobs for real-time data aggregation, features store for storage and reusability, this architecture improves both the accuracy and scalability of the customer journey modeling in a sequential recommendation system

\subsection{Batch and Real-Time Data Use Cases}
Our proposed data pipeline outputs the online and offline feature stores, expanding the capability of sequential recommendation models in the batched (offline model training, inference) and real-time (online learning, continuous training, inference) data use cases. 
% This eventually results in better recommendations and a deeper understanding of customer preferences.
% This step includes feature retrieval for training the recommendation model. During training, the features from both real-time and offline in-store data are retrieved from the feature store's offline store. This hybrid combination helps the model learn both the immediate user intent and batch in-store preferences.
 
% \subsection{Real-Time Inference} 
% Once the model is trained, it is deployed to serve real-time predictions. To enable real-time model inference, the necessary features were pushed to a cache, which allows low-latency access when generating recommendations for customers. During inference, the same features (or a subset) that were used for training are pulled from the online cache. The streaming job continuously feeds new real-time data into the system, and the model makes predictions based on both the new interactions and the stored features from the registry. The caching mechanism helps with online feature retrieval. It ensures that the features can be retrieved in milliseconds so that the model can access and utilize the most recent data quickly to deliver timely and relevant recommendations based on the latest available data.

\section{Models} \label{method}
As previously discussed, transformer-based sequential recommender systems cannot be directly applied to hybrid sequences by combining online and in-store user behaviors. 
In this session, we introduce an attention-based encoder that targets only in-store behaviors in the hybrid sequence without breaking the original transformer structure in the sequential recommender systems. 

\subsection{Store Transaction Encoder with Self-attention}

\begin{figure}
    \centering
    \includegraphics[width=\linewidth]{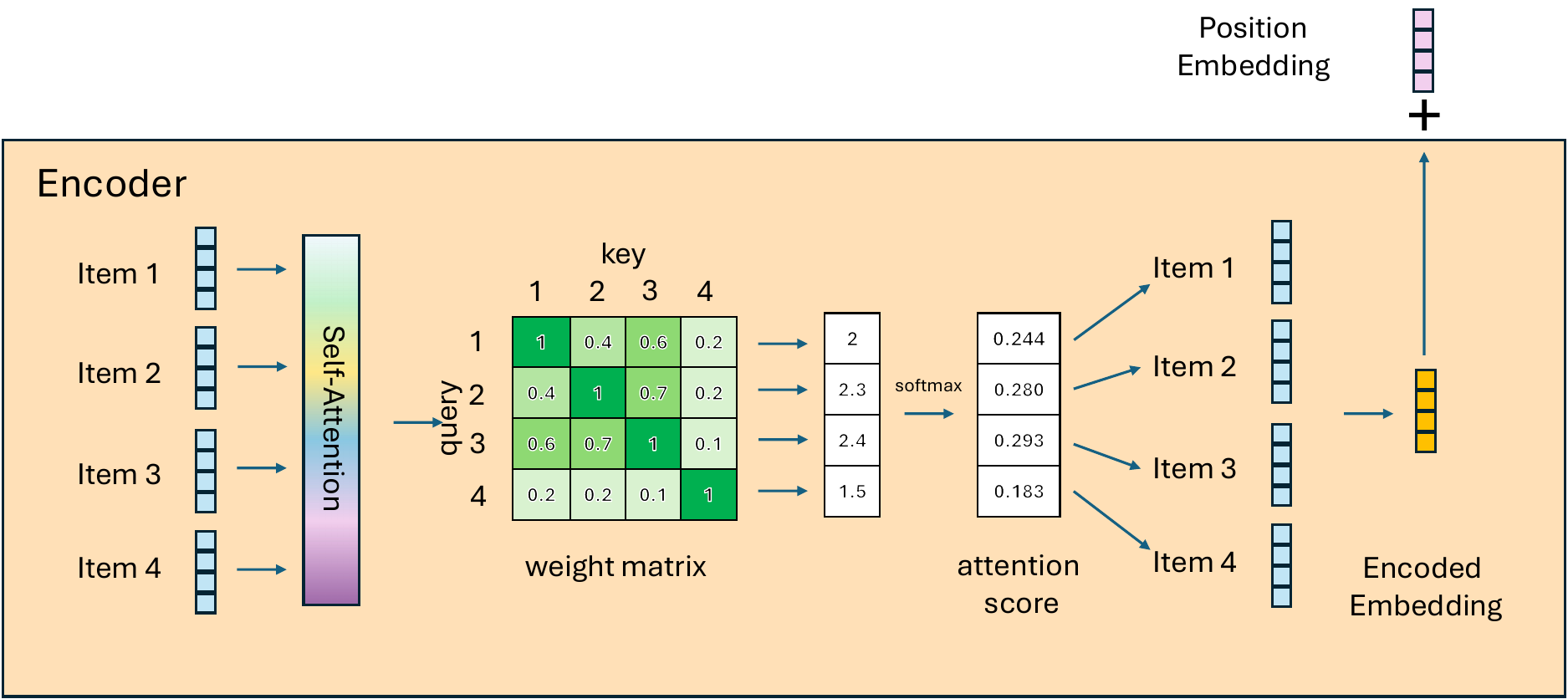}
    \caption{Store Transaction Encoder with Self-attention. The attention score is generated from the unnormalized self-attention weight matrix on the given item set, which is permutation invariant. The final encoded item set representation is a weighted sum of the item embedding.}
    \label{fig:encoder}
\end{figure}

Items from the same timestamp in in-store behaviors form a set that is permutation invariant. To enable transformer-based sequential recommenders on the hybrid user behavior sequence, we develop an encoder module to convert a set of items into a singular embedding vector instead of an embedding matrix. After this conversion, each timestamp will have only one input embedding vector, which fits the requirement of transformer-based recommender systems. 

We share the detailed design of the encoder in Figure~\ref{fig:encoder}. To meet the requirement on permutation invariant, we consider the self-attention mechanism to learn the self-attention weight matrix. 
Given a set of items from in-store behavior $s$ with the cardinality as $n$, we first convert these items into their item embeddings and form an embedding matrix $m_s \in \mathcal{R}^{|n|\times|d|}$. Then, we apply self-attention and compute the weight matrix $w_{attn} \in \mathcal{R}^{|n|\times|n|}$ by Equation~\ref{eq:attention}, where projection matrices $w_q, w_k \in \mathcal{R}^{|d|\times|d_a|}$.
\begin{equation}
    w_{attn} = QK^T, Q = m_s w_q, K = m_s w_k
    \label{eq:attention}
\end{equation}
The actual attention score $a_s \in \mathcal{R}^{|n|}$ is computed by a normalized row-wise summarization from the weight matrix $w_{attn}$, Equation~\ref{eq:score}, where $\mathbf{1}$ is a $n$-dimensional vector with all elements are $1$.
\begin{equation}
    a_s = \text{softmax}(w_{attn} \mathbf{1})
    \label{eq:score}
\end{equation}
Finally, the encoded representation of $s$ is computed by Equation~\ref{eq:sum} as a weighted sum. 
\begin{equation}
    v_s = a_s \cdot m_s
    \label{eq:sum}
\end{equation}

This design aims to represent the item set by weighing different shopping intents within the same item set.
The weight matrix $w_{attn}$ first measures the inter-item similarity. The row-wise aggregation collects the similarity statistics for each item and reflects the dominance of shopping intents in the item set -- if an item belongs to a dominant shopping intent, it will be similar to more items than those not in the dominant shopping intent. 
Thus, the final attention score vector approximates the underlying distribution of shopping intents on each item. 

\subsection{Training and Inference}
% Because the input sequence consists of a combination of individual items and item sets, the training and inference processes need several changes to enable the encoder in the transformer-based sequential recommender systems. 

During data preparation, we reserve a special token to represent the item set in the sequence and keep the actual item set as meta information. During training, when processing the special token, the encoder processes the item set from the meta information. The output is redirected to the transformer backbone for regular gradient back-propagation. 
We use the last position's output to predict the next item of interest. 
In many transformer-based sequential recommender systems, the cross-entropy loss is employed to optimize the classification of the target item out of the entire item cohort. 
We can still apply the same optimization practice over the entire model, and the gradient will be back-propagated to the encoder module for update. 
However, the special token should not be on the positive labels or the negative samples. 
During inference, the gradient back-propagation is stopped, and the special token should not be in the candidate set as we predict the future online user behavior.

\section{Evaluation} \label{eval}
In this section, we present our experiment setup and the analysis results. 
Our experiments are designed to answer the following research questions:
\begin{itemize}
    \item RQ1: Could in-store user behavior improve the prediction of future online user behavior? 
    \item RQ2: What is the influence of the store transaction encoder? 
    \item RQ3: How effective is the attention mechanism in the encoder? 
\end{itemize}

\subsection{Dataset}
We collect the dataset from a real-world e-commerce platform\footnote{Walmart.com} by anonymizing the user and item information. 
We use the hybrid data pipeline mentioned in Section~\ref{sec:data_pipeline} and sort the user's online and in-store behaviors by their timestamps in ascending order per user $u$. 
This real-world dataset has $58766$ users, $317457$ unique items, with at least $20$ behaviors (i.e., $|h_u| \leq 20$). More specifically, the average length of a user behavior $|h_u|$ is $33$, and $40\%$ of behaviors are in-store. 
To generate the training and evaluation labels for each user, we keep the last online behavior for evaluation and the second last online behavior for training.

\subsection{Baseline and Variants}
To show the effectiveness of our method, we consider SASRec~\cite{kang2018self} as the model backbone of transformer-based sequential recommendation models.
We also developed different variants of SASRec and our proposed model to study the research questions. 
\begin{itemize}
    \item $\text{SASRec}$: the original SASRec trained on the pure online behavior sequences. 
    \item $\text{BERT4Rec}$: the original BERT4Rec trained on the pure online behavior sequences. 
    \item $\text{SASRec}_{w/ store}$: the original SASRec trained on the hybrid behavior sequences. 
    \item $\text{SASRec}_{attnE}$: the modified SASRec with the self-attention-based encoder for the in-store behavior, trained on the hybrid behavior sequences. 
    \item $\text{SASRec}_{avgE}$: the modified SASRec with the encoder for the in-store behavior based on average-pooling instead of attention scores in Equation~\ref{eq:sum}, trained on the hybrid behavior sequences. 
\end{itemize}

\subsection{Implementation Details}
For the architecture in the default version of SASRec, we follow the implementation suggested by~\cite{kang2018self}.
For a fair comparison, we consider the suggestion from~\cite{klenitskiy2023turning} and leverage negative samples to compute the cross-entropy loss for all models. 
The optimizer is the Adam optimizer~\cite{kingma2014adam}.
The learning rate is 0.001, and the batch size is 512. The dropout rate of turning off neurons is 0.2.
The maximum sequence length is set to 90. 
We trained each model with 20 epochs and kept the best-performing model checkpoint for evaluation. 

\subsection{Evaluation Metrics}
For evaluation metrics, we adopt two common Top-N metrics, Hit Rate@10 and
NDCG@10, to evaluate recommendation performance as suggested by~\cite{kang2018self}.
During the evaluation, we keep the positive label for each user $u$ and randomly sample 100 negative item IDs (excluding the special token for online behaviors) to avoid heavy computation. 
Based on the rankings of these 101 items, we report the Hit@10 and NDCG@10.

\subsection{Recommendation Performance}
We summarize the model performance on the real-world dataset in Table~\ref{tab:results}. 
First of all, $\text{SASRec}$ outperforms $\text{BERT4Rec}$ after using sampled cross-entropy loss, which is aligned with the trend in ~\cite{klenitskiy2023turning}.
Further, we can see that compared with $\text{SASRec}$, all other models trained on the hybrid behavior sequences achieve better performance in both HitRate@10 and NDCG@10, which answers the first research question (RQ1). 
Comparing the models with encoders ($\text{SASRec}_{avgE}$, $\text{SASRec}_{attnE}$) to models without encoders ($\text{SASRec}_{w/ store}$), we can see that a general design of encoders on in-store user behaviors improve the learning. 
We guess this is because in-store user behaviors still share a certain level of independence compared with online user behaviors due to realistic considerations such as time, transportation, item in-store eligibility, etc. 
The encoder could serve as an adaptor to summarize in-store user behavior and fit the signals into the online user behaviors for future prediction (RQ2). 

Moreover, we can see that the model variant $\text{SASRec}_{attnE}$ with the attention mechanism encoder achieves even better performance compared with the average-pooling strategy. This indicates that in-store user behaviors also have different shopping intents, and the attention mechanism could learn this complexity better (RQ3).

\begin{table}[]
\centering
\caption{Recommendation Performance: . The best-performing method is boldfaced. The percentage of performance improvement compared with the baseline model $\text{SASRec}$ is provided per metric. }
\begin{tabular}{ccc}
\hline
  & HitRate@10 & NDCG@10 \\
\hline
$\text{BERT4Rec}$ &      0.301     &    0.197  \\
$\text{SASRec}$ &      0.387     &    0.240  \\
$\text{SASRec}_{w/ store}$ &     0.393 (+1.55\%)     &    0.246   (+2.50\%)  \\
$\text{SASRec}_{avgE}$ &      0.401 (+3.62\%)     &     0.255  (+6.25\%)  \\
$\text{SASRec}_{attnE}$ &     $\textbf{0.404}$   (+4.29\%)    &    $\textbf{0.258}$  (+7.50\%)  \\
\hline
\end{tabular}
\label{tab:results}
\end{table}

\section{Conclusion} \label{conclusion}
In this paper, we study the sequential recommender system with an increasingly important data application combining online and in-store user behaviors to predict future online behavior. 
We propose a hybrid data pipeline with configurable feature transformation, aggregation, and registry in different types of caches to support the life cycle of downstream machine learning models.
We propose an encoder compatible with transformer-based sequential recommender systems to model hybrid user behaviors better. 
The experiments on the real-world dataset demonstrate the effectiveness of the hybrid data pipeline and the encoder.

% \newpage
\bibliographystyle{IEEEtran}
\balance
\bibliography{main} 

\end{document}